# RD53A pixel module assembly and testing experience

**A. Petrukhin on behalf of the ATLAS ITk Pixel Collaboration**

*Center for Particle Physics Siegen, Experimentelle Teilchenphysik, Universität Siegen*
*E-mail*: petrukhin@hep.physik.uni-siegen.de

ABSTRACT: The entire tracking system of the ATLAS experiment will be replaced during the LHC Phase-II shutdown. A new silicon Inner Tracker (ITk) will contain five pixel layers equipped with new sensors and readout electronics capable to improve the tracking performance, cope with the high particle multiplicity and work in a high-luminosity environment. In order to standardize modules of the Inner Tracker, the idea of a common hybrid pixel module was introduced to be used in all regions of the detector. The common hybrid module assembly and testing techniques will be presented. The module construction, metrology and electrical testing results will be discussed. The work was done with the RD53A pixel modules – prototype modules towards the final ITk pixel detector.

## Contents

## 1. Introduction

The entire tracking system of the ATLAS experiment [1] will be replaced during the Large Hadron Collider (LHC) Phase-II shutdown. A new silicon Inner Tracker will contain five pixel layers equipped with new sensors and readout electronics capable to improve the tracking performance, cope with the high particle multiplicity and work in a high luminosity environment provided by the LHC. The total surface area of silicon covered by the new pixel detector will be ~14 m$^2$. About 10 000 pixel modules will be built for the Inner Tracker during 2 years of production. 22 different institutes are involved in the pixel module building and electrical testing. Each "bare" module consists of a silicon sensor divided into pixels and connected to the Front End (FE) readout chips by bump bonds with a high density of more than 35 000 contacts/cm$^2$. A quad module with 4 FE chips provides almost 320 000 readout channels after the hybridization [2]. The signal transmission from these readout channels as well as the power distribution for the modules is achieved using a flexible hybrid PCB. This is a custom-designed flexible circuit produced in industry. The sensor bias voltage of 100 V before and 600 V after irradiation is also distributed by the hybrid flex PCB. In order to standardize modules of the Inner Tracker, the idea of a common hybrid module was introduced to be used in different regions of the detector. One of the solutions for this concept is to use detachable pigtails, which are flexible printed circuits with different lengths, to connect the common flex hybrid in the different detector regions. Taking into account that about ten thousand pixel modules have to be assembled and working for years in harsh conditions, the choice of FE chips, sensors, the common flex hybrid design and the module assembly method is very important for the detector upgrade. Other very important parameters such as the amount of dead material in the sensor area, the power dissipation of the module and a possible mismatch in the coefficient of thermal expansion of different module materials are also considered.

In the following the pixel module assembly and electrical testing procedures are described, as implemented by the ATLAS group of the University of Siegen in close collaboration with other institutions.

## 2. Pixel module assembly

There are several steps in the construction of the pixel modules. At the very first step the custom bare module and the flex PCB have to be made and the parts should be fabricated in industry. All assembly procedures at the module building sites have to be performed in a clean room with a particle count equivalent to class ISO 7 with a controlled temperature of 20 ± 2°C and a relative humidity of 50 ± 5%. Module components arrive cleaned per industry standards and no additional cleaning is required at the assembly site. The assembling operator is always grounded to earth for electrostatic discharge protection in order to keep modules electrically safe. Before building of modules, the module assembly sites undergo a qualification process using glass and silicon dummies in order to show that a site is follows the common assembly procedure and to ensure a high yield with the RD53A quad modules.

Assembly starts with a visual inspection. The purpose of visual inspection is to identify any damage to the module like scratches, dirt, broken chips etc. A visual inspection is done by two methods: with a digital microscope and with a 190 Mpix high resolution camera. All surface defects are documented with pictures and stored into the production database (PDB).

The electrical reception test is done by measuring the current-versus-voltage (IV) curve with a probe station. This test verifies that the sensor was not damaged during the hybridization process or during the transportation. A bias voltage is applied between the sensor backside and one of the ground pads on the FE chip. The voltage is increased with 5 V steps. A leakage current is measured 10 times at each voltage step and averaged. A damaged module is expected to show a breakdown below -200 V. A typical IV curve can be seen in Figure 1.

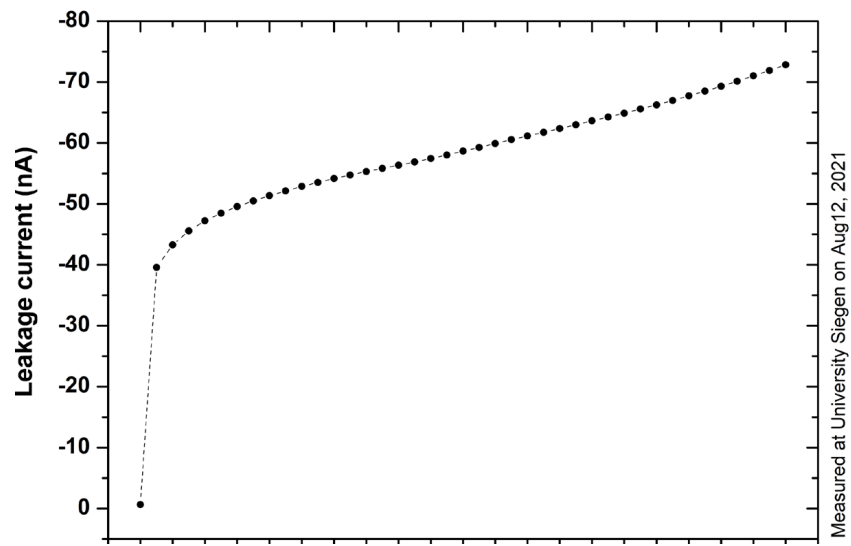

**Figure 1.** A typical IV curve of a bare RD53A module, measured using a probe station.

If the bare module shows a good IV characteristics, the hybrid gluing procedure is started. A preliminary step in this procedure is the calibration of the assembly tooling for the foreseen thickness of approx. 600 μm for thin FE chip modules. This step is necessary in order to get the correct glue thickness of 40 ± 25 μm. The module assembly tooling is shown in Figure 2. The gap between two parts of the tooling is calibrated with three precise adjustment screws on the flex PCB jig.

The adhesive used for the module attachment is Araldite 2011 [3]. It is a two-components epoxy glue, to be mixed right before the assembly. Any air bubbles in the glue should be avoided. The glue deposition is done with the stencil method. A small amount of glue is distributed on the flex surface through the stencil holes with a spatula. The stencil is shown in Figure 3 (left). After the glue is deposited, both jigs are put on top of each other and vacuum is applied. The curing time is approx. 24 hours. The glue must cover most of the flex area, especially beneath the wire bonding pads, to ensure good wire bonding and in four service pads which will be used for loading of the module to the support structure.

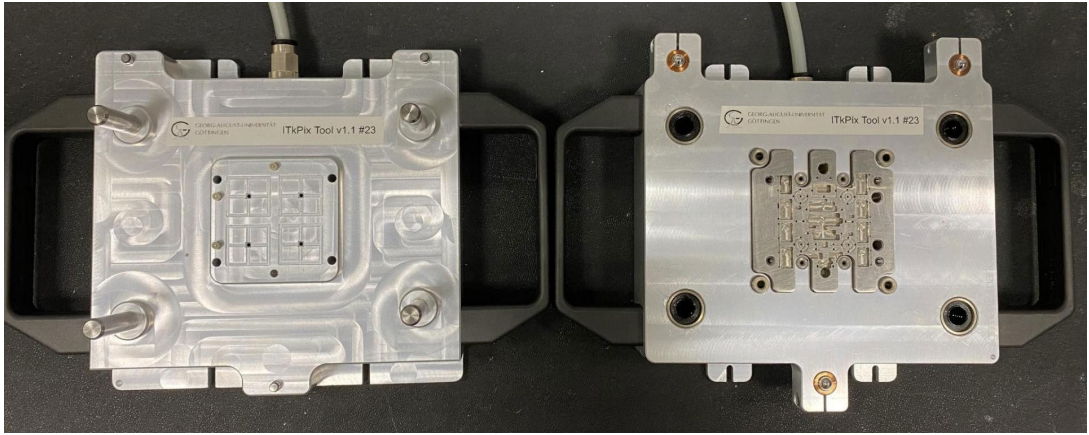

**Figure 2**. RD53A module assembly tooling with the bare-module jig (left) and the flex-PCB jig (right). The vacuum connections on top are also shown.

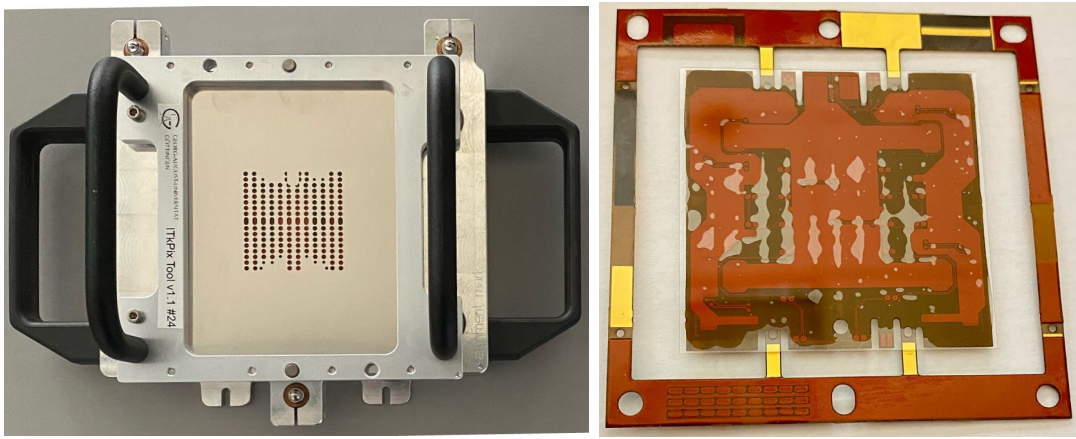

**Figure 3.** Left: stencil tooling placed on top of the flex jig. Right: glue distributed on a dummy glass module.

After the module is glued, the quality control (QC) procedure is used to verify the module attachment process. The visual inspection of the module should show that no glue seepage happened to the wire bonding pads and to HV connection hole. Once the module passed this criterion, metrology is performed by a non-contact method. The height of the module and the alignment in the *x-y* directions are then measured. The measurements are repeated with and without vacuum to estimate the flatness of the glued module. The glue thickness is calculated as the difference between the bare components and the assembled module. The module alignment is evaluated from the distance between the bare module and the flex edges on all four module sides. The maximum dimensions of the module should be inside of the module envelope defined in Ref. [4]. All measured data including the glue mass are uploaded to the PDB. The coordinate measuring machine and a typical flex PCB profile measured by this machine are shown in Figure 4.

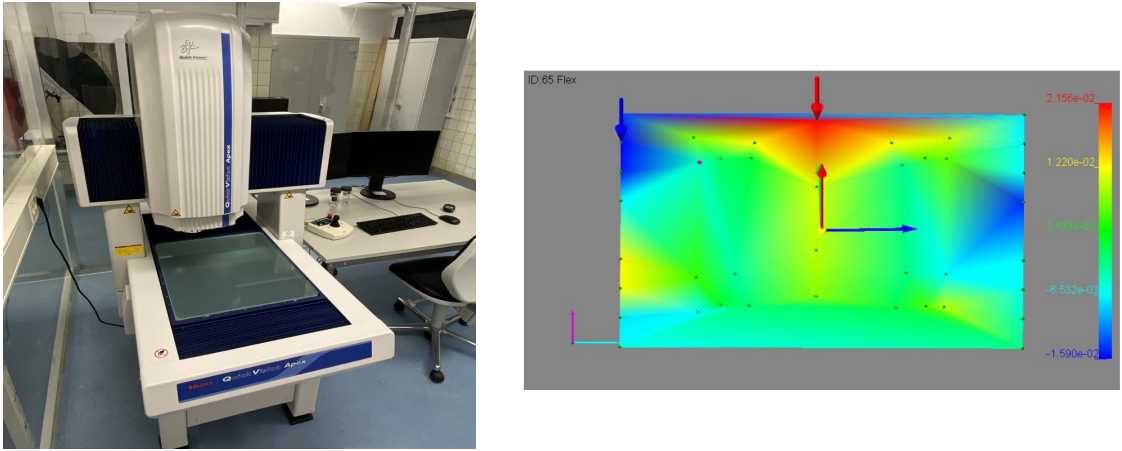

**Figure 4.** The coordinate measuring machine at the lab in Siegen (left) and a flex PCB profile (right).

Wire bonding is the next step in the pixel module building. It is done with an ultra-sonic wedge using a 25 μm diameter aluminium wire. In the case of RD53A modules, wire bonding was done at the University of Bonn as in 2020 there was no wire bonder available in our lab, yet. The QC of the wire bonding includes visual inspection to ensure that there are no shorts due to bad bonding. Another QC test is pull-testing. 8 dedicated wires per chip are pulled and the force needed to break the bonds is recorded. It should be greater than 5 g for all wires and 8 g or higher in average. 90% or more of breaks should be heel breaks. All built modules passed the criterion.

## 3. Electrical testing of modules

Once the module is assembled, its electrical functionality is checked. In Siegen, the module is placed inside a 3D-printed box with circulating dry air, cooled by a Peltier element and connected via custom-made adapters to the readout system. A requirement for the cooling unit is to keep the module temperature below 55°C during the electrical tests. Two readout systems were operated for the module testing. One is the YARR [5] system, based on the concept of moving intelligence from the Field-Programmable Gate Array (FPGA) firmware into the computer software. Another system is BDAQ [6], developed as a verification framework for readout chips of the RD53 family. It consists of custom and commercial hardware which is implemented to the BDAQ53 readout board, driven by an FPGA firmware and a Python-based data readout software. There are a few different types of readout tests which were used in the RD53A program. The basic functionality tests, which are used to check the digital and analog parts of the readout chips, are followed by the FE chip threshold tuning, the masking of noisy pixels and the threshold scan. The basic tests are mainly checking the wire bond connections while further tests allow to probe the quality of the modules. The occupancy maps of probing digital and analog parts are presented in Figure 5. After the basic functionality is proven, the ability of the module to register signals and transmit the data correctly are checked. This test verifies the quality of bump bond connectivity between the readout chips and the sensor. It is performed with a Strontium-90 source and a bias voltage of -100 V is applied to the sensor. Since the self-triggering mechanism was not implemented to the RD53A chips, the source scan was running for 48 hours in order to reach 50 hits/pixel. A climate chamber was used as a cooling unit for this test. The test setup and the source scan results are presented in Figure 6. One can see the pre-cooling system for the dry air, custom made passive cooling and adapters support block, and the source hood with the dry air circulating inside.

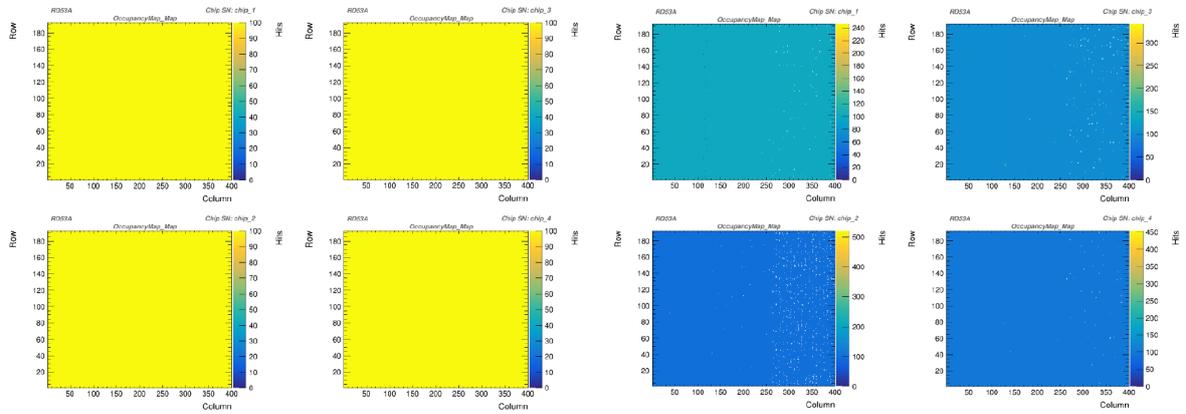

**Figure 5.** The occupancy maps from digital (left) and analog (right) parts.

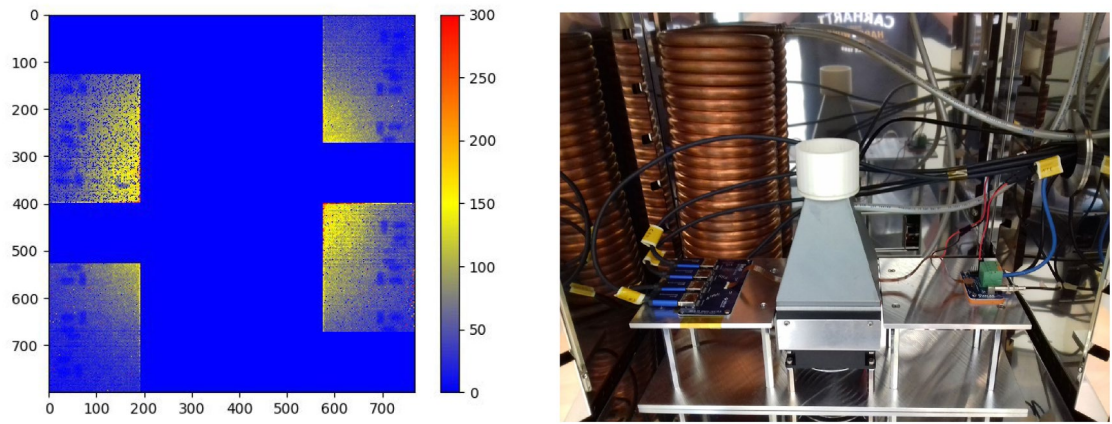

**Figure 6.** The number of hits collected during the source scan for linear and differential flavours of FE chips (left) and the test setup inside of the climate chamber (right).

The last test of the module is a thermal cycling. The module undergoes 10 thermal cycles between -45°C and +40°C. One additional cycle from -55°C to +60°C is performed. No corrosion, cracks or delamination should be observed during the visual inspection after all cycles. The thermal cycling procedure was done in a climate chamber with the module kept inside of the custom-made module carrier. The module was inside of the dedicated hood, with dry air circulating inside to avoid condensation at low temperature. The temperature was monitored by the climate chamber, a high-precision probe HygroFlex 5 [7], and an NTC thermistor located on the module PCB. The relative humidity results were provided by the climate chamber and the same high-precision probe located inside of the hood. The temperature was changing with a speed of 1°C per minute. The measurements of few thermal cycles are shown in Figure 7. One can see that the dew point was kept at 10°C lower than the module temperature.

After all the tests, 7 fully operational RD53A quad modules were packed into the vacuum sealing bags and shipped to CERN. Four of them were integrated into the Outer Barrel demonstrator being built in the framework of the RD53A program. All logistics was successfully performed with commercial antistatic boxes. Valuable experience was gained for further ITk pixel upgrade work.

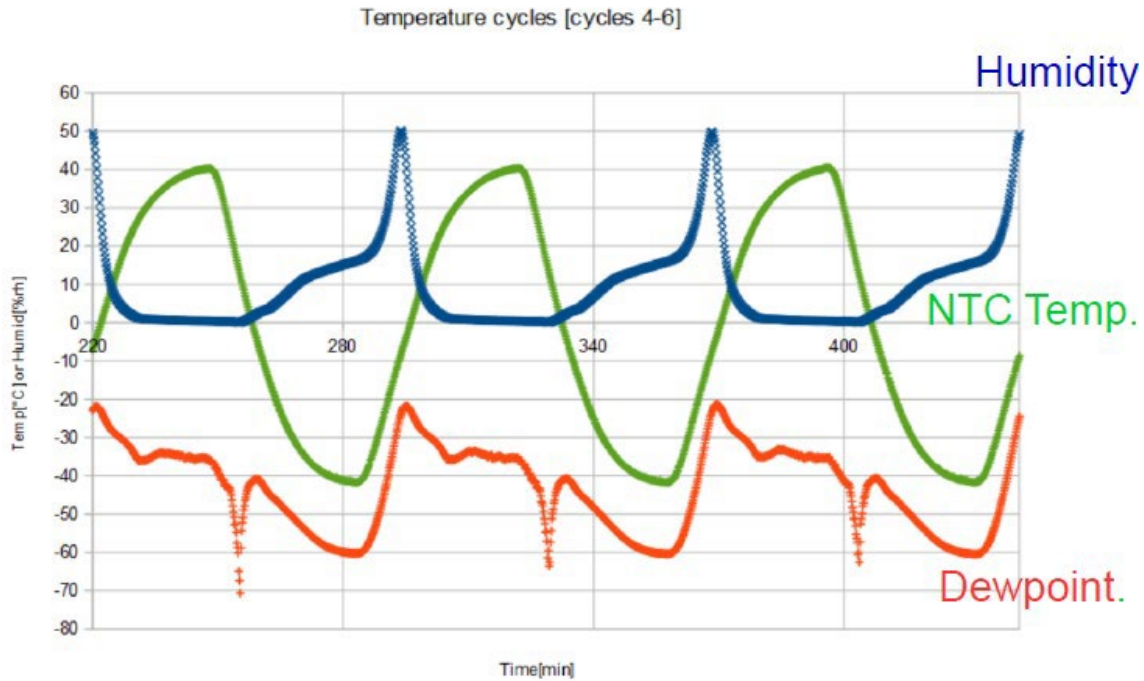

**Figure 7**. The results from three thermal cycles: relative humidity (blue dots), NTC temperature (green dots) and dew point (red dots).

## Acknowledgments

We gratefully acknowledge help from the ATLAS Bonn group, especially from Dr. Fabian Huegging, Wolfgang Dietsche and Susanne Zigann-Wack, in the module wire bonding and testing. Research is funded by German Federal Ministry of Education and Research and Heisenberg Program from the German Research Foundation.